\documentclass[]{aa}
\usepackage{graphicx}
\usepackage[]{natbib}
\bibpunct{(}{)}{;}{a}{}{,}
\def\simlt{\lower.5ex\hbox{$\; \buildrel < \over \sim \;$}}
\def\simgt{\lower.5ex\hbox{$\; \buildrel > \over \sim \;$}}

\begin{document}

   \msnr{h3240}

   \title{The Galactic Centre Source IRS~13E:  a Post-LBV Wolf-Rayet Colliding Wind Binary?}

   \author{R.F. Coker\inst{1} \and J.M. Pittard\inst{1} \and J.H. Kastner\inst{2}}

   \offprints{R.F. Coker, \\ \email{robc@ast.leeds.ac.uk}}

   \institute{Department of Physics and Astronomy, University of Leeds, 
              Leeds LS2 9JT  UK 
         \and
              Chester F. Carlson Centre for Imaging Science,  \\ Rochester Institute of Technology,
              54 Lomb Memorial Drive, Rochester, NY 14623 
             }

   \date{Received 2001 / Accepted 2001}

   \titlerunning{The GC Source IRS~13E}
   \authorrunning{Coker, Pittard, \& Kastner}

   \abstract{
IRS~13E is an infrared, mm and X-ray      
source in the Galactic Centre.
We present the first {\it Chandra} X-ray spectrum for IRS~13E and show that it
is consistent with a luminous and highly absorbed X-ray binary system.
Since
the X-ray luminosity is too large for a solitary star,
our interpretation is that of
an early-type long-period binary with strong colliding winds emission.
This naturally explains the observed X-ray spectrum and count rate
as well as its lack of significant short term variability.  Due to the
short lifetime of any nebula $0.2$~pc from the putative central
super-massive black-hole, we argue that the primary of IRS~13E
has exited the LBV phase in the last
few thousand years.
\keywords{X-rays: stars -- Stars: Wolf-Rayet -- Stars: individual: IRS~13E}
    }
\maketitle

\section{Introduction}

It is probable that Sgr A*, the compact, nonthermal radio source
at the Galactic Centre (GC) is a 2--3$\times10^6 M_{\sun}$ black-hole
\citep[for a recent review see][]{MF01}.  
Pervading the central parsec of the Milky Way
is a cluster of a few dozen HeI and
early-type stars \citep{SMBH90,GTKKT96}.
One of these stars, IRS~13E, has been identified as a Wolf-Rayet (WR) star with
spectral class WN10 \citep{Najarroetal97} and lies within IRS~13, a
compact HII region.  The IRS~13 complex, dominated by IRS~13E, has been
identified as a HeI, Pa-$\alpha$, [FeIII], and HeII line
source \citep{LKG93, LPFA95, Krabbeetal95, Stolovyetal99}.

Motivated by the conjecture \citep{CP00} that IRS~13E
is an X-ray binary, we present here an analysis of {\it Chandra} Advanced CCD
Imaging Spectrometer (ACIS) GC observations (obs ID=242) of this source.  Details
of the observations can be found in \citet{Baganoffetal00,Baganoffetal01b,Baganoffetal01}.
The X-ray source $\sim4\arcsec$ west-southwest of Sgr A* is the only 
source in the central parsec that has been
associated with a previously known stellar object, IRS~13E.
Since IRS~13E appears as an X-ray source while the other early-type
stars in the central parsec do not, IRS~13E must harbor a distinctive object.

Based on its lack of significant variability at all wavelengths and its
strong X-ray luminosity with characteristic $kT \simeq 1.0$ keV,
we argue that IRS~13E is most likely an early-type wide binary system
with the primary only recently having exited the luminous blue variable (LBV) phase of evolution.

\section{Source Identification}

As discussed in \citet{CP00}, we identify the X-ray source
$3\arcsec$ west\footnote{At 
a distance of $8.0$ kpc \citep{R93},
$1\arcsec \simeq 0.04$ pc.} and $1.5\arcsec$ south of Sgr A* with
the infrared source IRS~13E.  However, further identification is far from clear.
In K-band observations the IRS~13E complex is resolved into
3 components \citep{OEG99}:  IRS~13E1 $(m_K = 10.26)$, IRS~13E2 $(m_K = 10.27)$, and
IRS~13E3 $(m_K = 10.31)$ \citep[Fig.~8 of][]{PMMR01}.
However, these objects do not match up with mm observations, which show a different
set of $3$ components, some of which are extended \citep[Fig.~2 of][]{ZG98}.  
We previously \citep{CP00} identified the mm sources IRS~13E and IRS~13W with
the IR sources IRS~13E1 and IRS~13E2, respectively, but
careful astrometry shows that none of the K-band sources are coincident
with the mm sources (Maillard 2001; private communication).  This implies the mm
sources may not be stellar; the relative locations of the mm and IR sources open the possibility
that the radio sources are due to the colliding winds of a binary -- or even triple -- system.  
However, it appears that in agreement with \citet{PMMR01},
IRS~13E3 is slightly extended so its identification as a single stellar source is unclear.

\citet{PMMR01} identify IRS~13E3 as the
HeI source while \citet{Najarroetal97} identify IRS~13E1 as the HeI source.
However, NICMOS data, taken in March 1998, 
show (Stolovy 2001; private communication) that IRS~13E1
has very little (if any) excess emission at $1.87\mu$m while IRS~13E3
has some excess and IRS~13E2 has a strong excess.  Since a $1.87\mu$m
excess probably represents a blend of Pa-$\alpha$, HeI and HeII emission lines,
IRS~13E2 is very clearly the dominant Pa-$\alpha$ and HeI
source of the IRS~13E complex.

\citet{PMMR01} also report that their June 1998 observations show IRS~13E3 is 
slightly elongated and much dimmer
$(m_K = 11.73)$
than the other two components.  But the series of observations by
\citet{OEG99}, undertaken from 1992 to 1998, show that all three
components are of nearly equal magnitude and that IRS~13E3 (their source 21)
in particular is non-variable
at the level $0.2$~mag.  Further, K-band images taken in 1993 and 1994 \citep{Tamblynetal96}
which do not resolve the IRS~13E complex have $m_K = 9.1$~mag, fully consistent
with the 3 equal sources with $m_K = 10.3$~mag seen by \citet{OEG99}. 
The cause of this $\sim 1.5$ mag discrepancy is unclear.

Although not critical to our results, in this paper we will identify the
WN10 star with IRS~13E2 and the X-ray source with the IRS~13E complex as a whole.
The components of IRS~13E are separated from each
other by $\sim~0.1$--$0.2\arcsec$ in projection, so that, given
the $\sim 0.5\arcsec$ resolution of {\sl Chandra},
we lack sufficiently good positional references for
precise identification of the X-ray source.

\section{The X-ray Data and Spectral Fitting}

{\it Chandra} observed the GC in September 1999 and again in
October 2000 \citep{Baganoffetal00,Baganoffetal01b}.
Here we present detailed analysis of only the first epoch of observations.
We used standard data processing tools available from the Chandra X-ray
Center (CXC) as part of the CIAO v2.1 package\footnote{For details see
http://cxc.harvard.edu/ciao/index.html} to extract spectra of the X-ray
source and of adjacent background, and to determine the appropriate
response matrix and ancillary response (effective area) function for the
source extraction region. The source spectrum was extracted by binning
events lying within a $2.5\arcsec$ radius circle centered on IRS~13E;
the background spectrum was extracted from within three $2.5\arcsec$ radius
circles adjacent to and immediately north, south, and west of this source
region.  We avoided the region immediately to the east of IRS~13E, as it is
dominated by intense emission from Sgr A*.  Within the source region we
find 137 counts in 46 ksec net integration time. The background regions
contain $\simeq 45$ counts in an equivalent area, such that the
background-subtracted count rate is 0.002 cts s$^{-1}$.
Within the limited      
Poisson statistics, the X-ray flux
from IRS~13E appears to remain constant between the two epochs.
In addition, within the two $\sim 50$ ksec observations, IRS~13E is not seen
to vary.  

We fit the data with the MEKAL \citep{MGO85} model in
XSPEC\footnote{Distributed and maintained
by HEASARC}.  This is
a thermal plasma model and
assumes optically thin X-ray line and continuum emission.
We account for an external absorption column by simultaneous
application of the {\sl wabs} model.

In fitting the spectral X-ray emission, we vary three parameters:
the X-ray luminosity in the 0.2-10 keV band ($L_\mathrm{x}$), 
a characteristic temperature ($kT$) of the X-ray emitting gas, and 
the line-of-sight column density ($N_\mathrm{H}$).
The metallicity, $Z$, of the stars in the central parsec is not well 
known;  ISM gas phase abundances are roughly twice solar \citep{SCREH95} while
that of GC red super-giants is only solar \citep{CSB00,ramirezetal00}.  
The goodness of fit does not change significantly when the global 
abundance is varied so we have assumed solar metallicity for simplicity.

Spectral models can be fitted to data with 
low counts using the Cash statistic \citep{C79}.  However, a
background cannot be subtracted using this method.  We therefore 
employ the $\chi^2$ fit statistic to our background-subtracted data,
rebinned to a minimum of 10 cts per bin.

\section{ Results \& Discussion }

\begin{figure}
       \resizebox{\hsize}{!}{\rotatebox{-90}{\includegraphics{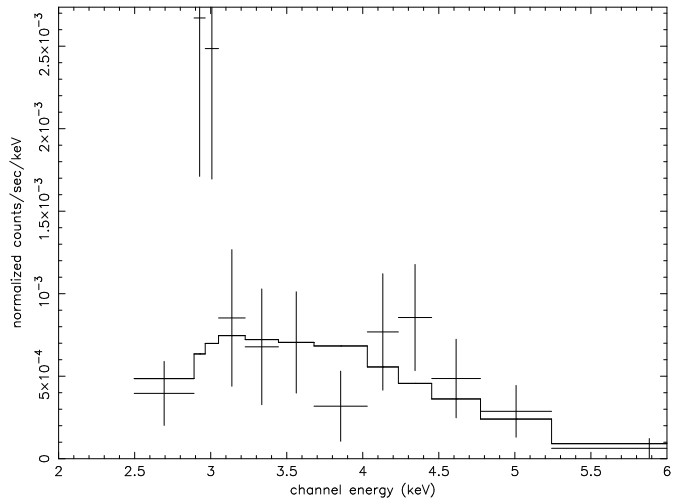}}}
      \caption[]{A plot of counts per second per keV versus energy for the data
        (crosses) and best-fitting model (solid line).  }
      \label{fig:1}
\end{figure}

Fig.~\ref{fig:1} shows the ACIS data along with the 
best-fitting model, which
yielded
$\chi^2_\nu \simeq 1.5$, formally a somewhat poor fit.
The best-fitting parameters are listed in Table~\ref{tab:fit}.
Inspection of Fig.~\ref{fig:1} shows that our model underestimates
the evident line emission at $\simeq 3$ keV and slightly overestimates the
emission at $\simeq 3.75$ keV.  Some of this may be due to poor
statistics but the strong emission near $3$ keV may also reflect a complex metal abundance.

\begin{table}
  \caption{Computed Model for IRS~13E and Arches A2}
  \label{tab:fit}
  \begin{tabular}{lll}
  \hline
 & IRS~13E  & Arches A2$^a$ \\
    \hline
 & \\[-7pt]
 $Z$              &1.0  & 2.6$^b$          \\
 $L_\mathrm{x}~(\mathrm{L}_{\sun})$               &7      &20       \\
 $kT$ (keV)                  &$1.0\pm0.4$ &$1.0\pm0.5$             \\
 $N_\mathrm{H} (10^{22}$ cm$^{-2})$         &$15\pm5$     &$12\pm2$      \\[3pt]
    \hline
  \end{tabular}

a) from \citet{Yusefzadehetal01b} \\
b) averaged Z over Si, S, Fe, Ar, and Ca  
\end{table}

Our fit suggests
$N_\mathrm{H}~\simeq~15\times~10^{22}$~cm$^{-2}$.  This is somewhat
larger than the $5$--$10\times~10^{22}$~cm$^{-2}$ found for Sgr A*
\citep{Baganoffetal00,Baganoffetal01b},
implying that there is additional substantial absorption close to the IRS~13E X-ray source.
If so, and if the size of the X-ray source is comparable to the $\simeq 0.1\arcsec$ diameter
region seen in mm observations, then the characteristic density of the gas surrounding IRS~13E is
consistent with a colliding wind binary (CWB) system \citep{SBP92} but
somewhat more dense than LBV ejecta \citep{Sewardetal01}.
However, some of our estimated column to the GC may be due to dust \citep{Baganoffetal01b}
since, unlike most of the massive stars in the central parsec, the IRS~13 complex is enshrouded by warm
dust \citep{M99}.  Similar large amounts of dust are seen around ``cocooned'' stars located in the 
Quintuplet Cluster that is
50 pc from Sgr A* in projection \citep{FMM99}; although these stars are probably WCd type stars,
this is not yet certain \citep{MSBFN01}.

The fitted characteristic temperature, $kT \simeq 1.0$ keV, is consistent with the
shocked winds of a CWB system such as a WR+O.  In contrast, an accretion
source with a compact companion such as in a massive X-ray binary (MXRB) system,
would typically have a harder spectrum \citep[{\sl e.g.},][]{Schlegaletal93}.
Also, a solitary massive O-star typically has a characteristic temperature of 
only $\sim~0.5$~keV \citep{CHS89}.
In the case of
a CWB, the characteristic temperature
represents a global average of the hot, shocked colliding winds,
while X-ray emission from a MXRB
probes the accretion disk.
On the other hand, the X-ray emission from LBVs is not well known; only
$\eta$ Car, a suspected binary whose primary is an LBV,
has a well-determined spectrum.
The emission from $\eta$ Car is a combination of hard ($kT \sim 5$ keV) compact emission
from the star and softer ($kT \sim 0.5$ keV) extended emission from
ejecta \citep{Sewardetal01}.
However, as a binary, the X-ray emission from $\eta$ Car 
may not be typical of solitary LBV stars.

The absorption-corrected X-ray luminosity of IRS~13E in the 0.2-10 keV band
is found
to be $L_\mathrm{x} \simeq 7$~L$_{\sun}$.  Although larger luminosities
are found for the brightest early-type binary systems
\citep[e.g. WR 140 and, assuming binarity, $\eta$ Car;][]
{Huchtetal94,CISP01}, $L_\mathrm{x} \simeq 7$~L$_{\sun}$ is still
brighter than the typical CWB system.  However, 
this is not wholly unexpected given the large
mass-loss rate and wind velocity of IRS~13E2 \citep{Najarroetal97}.  
In contrast, such an X-ray luminosity is a bit low for a
MXRB with an accreting compact source unless the
system has an unusually long period.
Additionally, the X-ray luminosity of a MXRB generally varies
on short time-scales.  Therefore, due to its low
luminosity, lack of variability, and relatively soft spectrum, 
it is somewhat unlikely that IRS~13E contains
a short
period MXRB.

The lack of variability also makes it unlikely that
the X-ray emission from IRS~13E is due to either a flaring proto-star or young stellar object (YSO).
The estimated intrinsic X-ray luminosity is also considerably higher
than any known YSOs \citep[{\sl e.g.},][]{Garmireetal00}.
However, \citet{Clenetetal01} show that the K-L colour
of the IRS~13 complex as a whole is consistent with the presence of YSOs.

Single massive stars seem to
obey the rough relation
$\mathrm{log_{10}} \left(L^\mathrm{ISM}_\mathrm{x}/L_\mathrm{bol}\right) = -7\pm1$ 
\citep{Pallavicinietal81},
although more recent work \citep{CG91,Moffatetal01} suggests the scatter may be more substantial
and the ratio may be half a dex larger.  $L^\mathrm{ISM}_\mathrm{x}$ is the X-ray luminosity
corrected for extinction
due to the ISM but not for intrinsic extinction \citep{WCDS98}.  If we assume that 
$N^\mathrm{ISM}_\mathrm{H} = 10^{23}$ cm$^{-2}$, then our
model for IRS~13E results in $\mathrm{log_{10}} 
\left(L^\mathrm{ISM}_\mathrm{x}/L_\mathrm{bol}\right) = -7$,
no greater than solitary X-ray sources.  CWB systems tend to have an enhanced X-ray
to bolometric
luminosity ratio by a factor of a few compared to solitary stars, but there is considerable
variation.
Possible contamination and large error bars in determining 
$\mathrm{log_{10}}
\left(L^\mathrm{ISM}_\mathrm{x}/L_\mathrm{bol}\right)$ for IRS~13E2 make it
difficult to draw any conclusions concerning binarity using this ratio.
For example, it may be that very little of the extinction is intrinsic to the IRS~13E system (see below);
this would result in a substantially larger $L^\mathrm{ISM}_\mathrm{x}$.

We must caution that given the low signal-to-noise of the data, the 
best-fitting parameters are not very well constrained.  For example, forcing 
$kT = 5$ keV and varying only $N_\mathrm{H}$ and $L_\mathrm{x}$
results in a fit with $\chi_\nu^2 = 1.9$.  However, based on observations
of other objects in the GC it is probable that 
$N^\mathrm{ISM}_\mathrm{H} \simgt 10^{23}$ cm$^{-2}$.  Given this additional
constraint, we can say that $L_\mathrm{x} \simgt 1 \mathrm{L}_{\sun}$
and $kT \simlt 1.8$ keV.

Also shown in Table~\ref{tab:fit} are the results for the bright
soft component of source A2 of the Arches
cluster located $\simeq 25$ pc from Sgr A* \citep{Yusefzadehetal01,Yusefzadehetal01b}.
The fit to this source implies a higher than solar metallicity
for a range of metals.
Although we assume solar metallicity for IRS~13E,
the peaks near $3$ and $4.25$ keV in Fig.~\ref{fig:1} are
possible indicators of high S and Ar
content.  Enhanced abundances for these elements lead to a
slightly better fit ($\chi_\nu^2 = 1.3$) but due to the low
number of counts in our spectrum, we do not attribute much
significance to this.
Our column density, characteristic temperature, and X-ray luminosity
are close to that for A2, which is also coincident with
a known IR stellar source, suggesting the objects are of similar nature.

In short, even given the large uncertainties,
the X-ray luminosity and characteristic temperature of IRS~13E do not favour
a single object (O-star, WR star, LBV, or YSO).  Also, the lack of variability is
inconsistent with a short-period MXRB.  On the other hand, all of the characteristics
of IRS~13E are fully consistent with a long-period CWB system.

\section{Evidence for a Recent Post-LBV System}

Although many significant details of massive star evolution
remain to be addressed, 
the present picture of massive star evolution is that stars with
zero-age main sequence (ZAMS) mass $\simgt 25~\mathrm{M}_{\sun}$ pass through the WR stage.
The evolutionary sequence for the more massive stars ($M_\mathrm{ZAMS} \simgt 40~\mathrm{M}_{\sun}$)
is thought to be \citep{Wa89,Langeretal94}:
\begin{eqnarray}\nonumber
\mathrm{O\rightarrow \mbox{H-poor}~WN \rightarrow LBV \rightarrow \mbox{H-free}~WN \rightarrow WC \rightarrow SN } \;.
\end{eqnarray}
Most stellar evolution models suggest it takes more than
$5\times10^5$ yrs for a massive star
to evolve through the entire WR sequence,
with the LBV phase taking more than $10^4$ yrs \citep{SC96}.
While in the LBV phase, massive stars are thought to repeatedly move back
and forth across the HR diagram, being bluer in quiescence and redder
during major eruptions.  More than $70\%$ of the LBV phase is spent in quiescence \citep{vG01}.
However, some evolutionary models suggest that
extremely massive
stars ($M_\mathrm{ZAMS} \simgt 60 \mathrm{M}_{\sun}$), 
particularly those with high metallicity or large rotation
rates \citep{MM00}, may never
become true LBVs \citep{CSHS95} or even WRs \citep{Mowlavietal98}.  

Since IRS~13E2, the presumed HeI line source (see \S 2), 
is spectrally a late-type WN star \citep{H01}, it can be
either a pre- or post-LBV object.  As chemically the source
has a large He to H ratio \citep{Najarroetal97}, IRS~13E2 is probably a post-LBV object.
There are thought to be a minimum of 
$6$ WC\footnote{However, \citet{PMMR01} claim some or even all of these
are actually not HeI stars and thus not likely WRs.} stars in the
central cluster \citep{H01} so at least some cluster members, presumably those with
the largest ZAMS mass if the cluster is coeval,
have already gone through
the LBV phase.  This is particularly noteworthy since the ZAMS mass of
IRS~13E2 is thought to be as high as $\simeq~120~$M$_{\sun}$ \citep{SSMM92}, implying
that the ZAMS mass of the 6 WC stars is even higher.  Some of the HeI stars in the central cluster
are comparatively high in H; they may be pre-cursor LBVs.  If the statistics
of LBVs versus WRs in the LMC is comparable to that in the GC, one would expect $\sim 1$
LBV out of a population of a few dozen WR stars \citep{HD94,P97}.

The H-poor WN stars, being pre-LBV, are more massive, more luminous and larger
than the H-free WN stars.  As the terminal velocity of radiatively
driven winds from early-type stars is in general correlated with their
surface gravity, the H-poor WN stars should have slower
winds and narrower lines than the H-free WN stars \citep[see, {\sl e.g.},][]{SSM96,Co99}.  Thus, the $7$ 
narrow-line stars that \citet{PMMR01}
classify as LBVs could be H-poor WN stars while the broad-line stars,
including IRS~13E2,
could be H-free WN stars.  This is consistent with He/H measurements
\citep{Najarroetal97} which correlate narrow-line stars with low He/H ($\simlt 3$)
and broad-line stars with large He/H ($\simgt 100$).  
While LBVs generally also have He/H of less than a few \citep{CSHS95},
it seems
unlikely, even in the peculiar environment of the GC, that $7$
LBVs would exist at the same time in the same cluster.
Comparison with WR 122, a star still shrouded in ejecta and thought to be just post-LBV,
might be useful.

Models of the bolometric luminosity and effective stellar temperature 
of IRS~13E2 ($M_\mathrm{bol} = -11.2$ mag
and $T_\mathrm{eff} \simeq 29$ kK) place it on the left edge of the
Humphreys-Davidson (HD) line, the location on the HR diagram where LBVs are thought to exist.
This temperature, as well as the temperature of the other early-type
stars in the central parsec, is somewhat colder than the lower limit
of traditional WR stars \citep[30kK;][]{W89}, but, due to the
high extinction, determining effective temperature
in the GC is notoriously difficult \citep{DCSB99}.
At present only a few central stars have estimated $M_\mathrm{bol}$
and $T_\mathrm{eff}$ values; within the relatively large errors, all
fall below or are near the HD line.  The
effect of metallicity on the HD line is unclear but models suggest that high
metallicity moves LBVs to lower $L_\mathrm{bol}$ and $T_\mathrm{eff}$ \citep{SC96}.
In the future, observations of CO absorption 
at $2.3 \mu$m and H$_\mathrm{2}$O absorption at
$1.9 \mu$m may better determine $T_\mathrm{eff}$ and $M_\mathrm{bol}$ for
IRS~13E2 and the other early-type stars in the GC \citep{BRS99}.

An LBV is often variable across many time-scales and
frequencies.  Although IRS~13E2 may be marginally variable at the
level of $\simlt 0.1$ mag in K-band \citep{OEG99}, 
it is not significantly variable in the mm \citep{ZG98} nor
apparently in the X-ray \citep{Baganoffetal01b}.
This suggests it has ended its LBV phase, although it may also be merely quiescent.
Since it takes $\simeq 3\times10^6$ years for a $100$ M$_{\sun}$ star to proceed
from the ZAMS through to the final WR phase \citep{Langeretal94}, this gives an approximate
age of the central cluster.

The high terminal wind velocity ($\simeq 1000$ km s$^{-1}$) and 
lack of short-term variability suggests that IRS~13E2 has ended
its LBV phase.  
Tidal disruption by Sgr A* means that any nebula surrounding IRS~13E2 is likely to be short lived
after the star becomes a broad-line H-free WN star.
In addition, since LBV terminal wind velocities can be lower 
than $\sim 200$~km~s$^{-1}$ \citep{NHS97,P97},
any GC LBV-spawned nebula will be distorted by stellar motion alone and thus would not
appear as symmetrical as, {\sl e.g.}, the Homunculus around $\eta$ Car.

\subsection{Lifetime of an LBV Nebula in the GC}

We can crudely estimate the ``half-life'' of an LBV
shell (that is, how long before the two sides of the shell are differentially stretched
by a factor of two if they are on separate circular Keplerian orbits) by
\begin{equation}\label{eq:1}
t \simeq {{2 D}\over{v_i - v_o}} \;,
\end{equation}    
where $D$ is the diameter of the shell and 
$v_i$ and $v_o$ are the circular Keplerian velocity at the inner and outer
edge of the shell, respectively.
Assuming the object is near enough to Sgr A* so that the
gravitational potential of the super-massive black-hole is dominant \citep[$\simlt 3$ pc;][]{GTKKT96},
the circular Keplerian velocity is:
\begin{equation}\label{eq:2}
v = \sqrt{{{GM}\over{R}}} = 100~R^{-1/2} \; \mathrm{km~s}^{-1}\;,
\end{equation}    
where $R$ is in pc.  Next, assuming $D \ll R$,
one finds that $t$ is independent of $D$:
\begin{equation}\label{eq:3}
t \simeq 4\times10^4 R^{3/2} \; \mathrm{yrs} \;.
\end{equation}  
Thus, for IRS~13E2, any LBV shell is likely to exist for only $\sim 2000$
yrs after the star's final molting.  Since the circular orbital period of an object 
within a few parsecs of Sgr A* is $6\times10^4 R^{3/2}$ yrs, the lifetime of any
nebula will be less than an orbit, regardless of $R$.

If the differential speed between the gas in the Bar and
IRS~13E is a few tens of km s$^{-1}$, a few thousand years is also the time-scale
needed for IRS~13E to move from the centre of the mini-cavity to its present
location \citep[see Fig.~1 in][]{CP00}.  Since the mm proper motion of the IRS~13E complex
points back towards the mini-cavity \citep{ZG99}
this raises the possibility of an association between the two:
{\sl e.g.}, did IRS~13E carve out the mini-cavity as it traversed 
the Bar?

Thus, there is strong evidence from the width of the He line at $2.058\mu$m
and the He/H ratio that IRS~13E2 has recently finished its LBV phase
and is currently a broad-line H-free WN star.  The presence of LBV ejecta
may also explain the high absorbing column and if true would place a
strong constraint on the time elapsed since becoming a H-free WN star.

\section{An Extended Binary System}

In order for IRS~13E to be such a strong X-ray source while
not being variable on time-scales of days or a year,
it is likely to be a wide binary with a period significantly longer than a year.  
IRS~13E1, a possible massive dwarf O-star, is approximately $0.15\arcsec$
from 13E2 and is a possible companion since much of the mm emission lies between
these two sources \citep{ZG98,ZG99}.  IRS~13E3 is $0.2\arcsec$ distant
in projection and, although it has weak Pa-$\alpha$/HeI emission so that it
may be a WR-star in its own right, there is no mm emission directly between
IRS~13E3 and IRS~13E2.  This implies that even if the three sources compose
a triple system, IRS~13E3 is too far away to produce significant emission
due to colliding winds.

A very crude estimate of the intrinsic column in a CWB is
\citep{U92,SBP92}
\begin{equation}\label{eq:usov}
N^\mathrm{INT}_\mathrm{H} \simeq 5\times10^{22} { {\dot{M}} \over { d v } } ~ \; \mathrm{cm}^{-2} \;,
\end{equation}
where the mass-loss rate $\dot{M}$ is in units of $10^{-5}~$M$_{\sun}~\mathrm{yr}^{-1}$,
the separation $d$ is in units of $10^{13}~\mathrm{cm}$ and the
wind velocity $v$ is in units of $1000~\mathrm{km~s}^{-1}$.
For sensible values of $\dot{M}$ and $v$, if any significant 
fraction of our estimated column is intrinsic, $d$
must be less than $\sim 100$ AU.
But the separation between IRS~13E2 and IRS~13E1 corresponds to about 700 AU
in projection, which, for a combined $70 \mathrm{M}_{\sun}$ system, implies
an orbital period of $\simgt 2000$ years.   At this separation the CWB
shocks are likely to be largely adiabatic and would emit only weakly in the X-ray 
\citep[$L_\mathrm{x} \propto {\dot{M}}^2 v /d$;][]{SBP92} unless the winds
are particularly dense.

However, on the plane of the sky, the IRS~13E complex is part of the
Bar in the center of the mini-spiral.  Thus, if the complex is
located within or behind the dense gas
of the Bar,
it is possible that much of our estimated column is due to the Bar
(which does not lie in front of Sgr A*) rather than the complex itself.
In addition, the extinction towards the GC is known to be very patchy
on small scales.
This suggests that IRS~13E1 and IRS~13E2 may still be companions, but
it would also imply that the LBV nebula has at least partially dispersed.

The X-ray luminosities of CWB systems vary with time as well as depending
on mass loss rates, wind velocities, and orbital separation.  Models of
WR 147, an extended CWB with a period of at least a few thousand years, predict 
$L_\mathrm{x} \sim 0.1 \mathrm{L}_{\sun}$ \citep{Pittardetal01},
However, the mass-loss rate of IRS~13E2 is estimated to be more than ten times that of 
the WR 147 primary.  Thus the X-ray luminosity of IRS~13E is consistent with
a CWB.

The proper motion of the mm sources in the IRS~13E complex have been
measured to be $\sim 250$ km s$^{-1}$ \citep{ZG98,ZG99}.  This is
consistent with the interpretation that they are
high-velocity knots within surrounding stellar ejecta.  
Observations show that at least one of the mm sources has a spectral index
suggestive of emission from ejected circumstellar material from
a stellar envelope.

In summary, the heavily absorbed X-ray spectrum of IRS~13E best matches a long period CWB system
whose primary has recently exited its LBV phase.
Long term monitoring at mm and K-band as well as long-integration 
{\it Chandra} observations will help determine precisely what type of 
binary is contained in the IRS~13E system.

\begin{acknowledgements}
This work was supported by PPARC and has made use of
NASA's Astrophysics Data System Abstract Service.  We
gratefully acknowledge helpful discussions with R. Oudmaijer, M. Hoare, 
S. Stolovy, and J.-P. Maillard.
\end{acknowledgements}

\bibliographystyle{apj}
\bibliography{references}

\end{document}